\documentclass[sigconf]{acmart}
\usepackage{algorithm}
\usepackage{algorithmic}
\usepackage{enumitem}
\setlist[itemize]{leftmargin=15pt}
\usepackage{titlesec}
\titlespacing*{\subsection}{3pt}{6pt}{0pt}
\titlespacing*{\subsubsection}{3pt}{4pt}{0pt}
\titlespacing*{\paragraph}{3pt}{3pt}{3pt}
\AtBeginDocument{%
  }
\setcopyright{acmlicensed}
\copyrightyear{2025}
\acmYear{2025}
\acmDOI{}
\acmConference[TSMO '25]{The 31st ACM SIGKDD Conference on Knowledge Discovery and Data Mining: Workshop on Two-sided Marketplace Optimization}{August 4, 2025}{Toronto, Canada}
\acmBooktitle{Proceedings of the 31st ACM SIGKDD Conference on Knowledge Discovery and Data Mining Workshop on Two-sided Marketplace Optimization: Search, Discovery, Matching, Pricing \& Growth (TSMO '25), August 4, 2025, Toronto, Canada}
\acmISBN{}
\begin{document}
\title{Counterfactual Reciprocal Recommender Systems for User-to-User Matching}
\author{Kazuki Kawamura}
\affiliation{%
  \institution{Sony Group Corporation}
  \country{Japan}
}
\email{Kazuki.Kawamura@sony.com}
\author{Takuma Udagawa}
\affiliation{%
  \institution{Sony Group Corporation}
  \country{Japan}
}
\email{Takuma.Udagawa@sony.com}
\author{Kei Tateno}
\affiliation{%
  \institution{Sony Group Corporation}
  \country{Japan}
}
\email{Kei.Tateno@sony.com}
\renewcommand{\shortauthors}{Kawamura et al.}
\begin{abstract}
Reciprocal recommender systems (RRS) in dating, gaming, and talent platforms require mutual acceptance for a match. Logged data, however, over-represents popular profiles due to past exposure policies, creating feedback loops that skew learning and fairness. We introduce Counterfactual Reciprocal Recommender Systems (CFRR), a causal framework to mitigate this bias. CFRR uses inverse propensity scored, self-normalized objectives. Experiments show CFRR improves NDCG@10 by up to 3.5\% (e.g., from 0.459 to 0.475 on DBLP, from 0.299 to 0.307 on Synthetic), increases long-tail user coverage by up to 51\% (from 0.504 to 0.763 on Synthetic), and reduces Gini exposure inequality by up to 24\% (from 0.708 to 0.535 on Synthetic). CFRR offers a promising approach for more accurate and fair user-to-user matching.
\end{abstract}
\maketitle

\section{Introduction}

Digital platforms increasingly rely on \emph{user-to-user} recommendation. For instance, online dating services have become highly prevalent, and similar matching is pervasive in multiplayer gaming, talent marketplaces, social audio, and peer-learning communities~\cite{potts2018reciprocal}. In these settings, a recommendation is successful only when \emph{both} parties accept it, making reciprocal recommender systems (RRS) fundamentally different from classic item recommenders~\cite{palomares2021}. The bilateral nature magnifies ethical stakes: a low-exposure user never even appears as a candidate partner, while popular or photogenic users absorb an outsized share of attention~\cite{celdir2024msom, bruch2018sciadv}.

The root cause is \textbf{exposure bias}. What the system shows today constrains the data available tomorrow, so learning on \emph{policy-filtered} logs reinforces the same skew---a self-perpetuating \emph{feedback loop} that widens inequality over time~\cite{chen2021feedbackloop, gao2024feedback, jannach2022popbias}. Longitudinal studies confirm that popularity bias inflates recommendation accuracy metrics offline while harming user welfare online~\cite{pisani2025exposure, chaney2018confounding, abdollahpouri2021usercentered}. Consequently, offline evaluation and model selection become unreliable, and fairness objectives such as two-sided envy-freeness are systematically violated~\cite{joseph2023twosided}.

Causal inference provides a principled remedy. Inverse Propensity Scoring (IPS) treats every historical recommendation as an intervention and re-weights each logged interaction by the inverse of its display probability, yielding a consistent estimator of the true risk~\cite{schnabel2016}.
Unfortunately, real propensities can be tiny, producing extreme weights and high variance; self-normalized IPS (SNIPS) \cite{swaminathan2015self} reduces variance but introduces a small finite-sample bias. Stabilized doubly robust (StableDR) estimators \cite{li2023stableDR} further aim to control variance and improve robustness when propensities are small, typically for \emph{item} recommendation.
Moreover, existing RRS work often overlooks challenges like bilateral exposure, pair-level propensities, and critical implementation details. The latter include consistently normalizing objectives over observed pairs (which SNIPS inherently addresses \cite{swaminathan2015self}) and effectively managing variance in such objectives.

We fill this gap with \emph{Counterfactual Reciprocal Recommender Systems} (CFRR). CFRR first estimates pair-level propensities to correct exposure bias. Then, it optimizes a self-normalized IPS (SNIPS) objective that provides stable training even with extreme weights. The framework naturally supports variance-control techniques---weight truncation and doubly robust augmentation---making it robust enough for industrial deployments where traffic is highly imbalanced. Extensive synthetic experiments demonstrate that, relative to strong RRS baselines, CFRR improves NDCG@10 by up to approximately 3.5\% (e.g., on DBLP, from 0.459 to 0.475), widens user coverage by up to approximately 51\% (on Synthetic, from 0.504 to 0.763), and trims Gini exposure inequality by up to approximately 24\% (on Synthetic, from 0.708 to 0.535), even when propensities are noisy.

\newpage
\noindent
\textbf{Our contributions are three-fold:}
\begin{itemize}
  \item \textbf{Bias-consistent and Stable Objective.} User-to-user matching as counterfactual estimation with a consistent IPS loss (correct propensities assumed). Our SNIPS-based CFRR objective provides a stable training framework that effectively handles extreme weights common in reciprocal matching scenarios.
  \item \textbf{Variance-Aware Learning for RRS.} CFRR integrates self-normalization across the entire objective and incorporates established variance reduction techniques like weight truncation and a carefully formulated doubly robust augmentation, adapted for the reciprocal matching context to mitigate variance from extreme weights.
  \item \textbf{Empirical Validation.} Controlled simulations and real-world dataset experiments confirm that CFRR can simultaneously boost ranking quality, expand long-tail exposure, and counter the feedback loop that otherwise amplifies popularity bias.
\end{itemize}

Despite its promising performance, deploying CFRR in a real system raises several additional considerations. We discuss in later sections how to handle extremely large user bases (positivity in practice), the impact of propensity model misspecification, and how CFRR can be extended to other two-sided platforms and fairness objectives.

\section{Related Work}
\label{sec:related_work}

This section reviews existing user-to-user matching approaches, highlighting limitations of methods ignoring selection bias. We then discuss fairness, two-sided constraints crucial in RRS, and situate our work concerning selection bias correction.

\subsection{Optimization-Based Methods}
\paragraph{Stable Matching.}
The classic Gale--Shapley algorithm \cite{gale1962college} for stable matching, initially for the stable marriage problem, guarantees stability (no unmatched pair mutually prefers each other). It is broadly applied in markets like college admissions \cite{roth1984evolution} and medical residencies. Yet, traditional algorithms assume complete preference orderings, often impractical for large-scale systems. They may also favor one market side, and direct application to dynamic platforms is challenging. Recent work explores scalability for large-scale RRS via parallel/mini-batch computations \cite{nakada2024parallel}.

\paragraph{Maximum Weight Matching.}
Other optimization methods model user pairs \((u,v)\) with a score, seeking to maximize total matched pair scores. This is often formulated as maximum weight matching (MWM) \cite{edmonds1972theoretical} or minimum-cost flow \cite{orlin1997polynomial}. They find global optima under capacity constraints (e.g., one match per user). However, MWM doesn't guarantee bilateral satisfaction or address score estimation bias from data, which can yield suboptimal/unfair outcomes.

\subsection{Machine Learning-Based Methods}
\paragraph{Collaborative Filtering.}
Techniques like collaborative filtering, originally for item recommendation, can be adapted to user-user matching by treating users as items \cite{pizzato2013recommending,xia2015reciprocal}. They predict user-user preferences from past interactions or profile similarities. However, sparse user-user data and unaddressed interaction biases in standard CF can yield poor generalization and reinforce popularity bias.

\paragraph{Reciprocal Recommenders.}
Reciprocal recommender systems (RRS) specifically model mutual preferences. They often learn distinct user representations (initiator/receiver) or use architectures for two-way dynamics \cite{xia2015reciprocal, neve2019latent}. Recent deep learning, like GNNs, better models complex RRS relations \cite{chang2022graphrr}. Though expressive for reciprocity, many assume observed interactions represent true preferences. This is often violated by selection/exposure biases from past policies \cite{schnabel2016}. Recent works offer new causal RRS formulations and metrics, noting unique domain challenges \cite{yang2024crrs}.

\subsection{Fairness and Two-Sided Constraints}
Reciprocal fairness is multifaceted, beyond relevance to equitable exposure/opportunity. Early work explored Rawlsian fairness \cite{joseph2016rawls} and Nash Social Welfare balancing \cite{rosenfeld2018fair}. Other research targeted equitable group exposure \cite{suhr2019twosided}. However, many such methods presume unbiased preference/compatibility estimates. Fairness/bias correction interplay is critical; fairness on biased data may not yield truly fair outcomes. Recent literature investigates new fairness definitions/algorithms for reciprocal settings (e.g., fair representation, mitigating discrimination \cite{tomita2024fairrec}). Effectively integrating robust bias correction with fairness objectives remains a challenge.

\subsection{Selection Bias in User-to-User Matching}
Selection bias plagues most RRS: training data isn't random, heavily influenced by past exposure policies. This creates feedback loops skewing data toward popular users \cite{chen2021feedbackloop}.
Causal inference offers a principled remedy. IPS and related estimators reweight/correct data for a balanced distribution (surveyed by \cite{gao2022causal}). Propensity score estimation is challenging; errors impact performance \cite{dudik2014doubly}. Applying these to RRS adds complexities: pair-level propensities (depending on both users) and bilateral outcomes. While some works apply causal reasoning to RRS (e.g., bilateral interventions \cite{yang2024crrs}, reciprocal link debiasing \cite{yang2024crrs}), a comprehensive framework for robust pair-level propensities, self-normalization, and practical variance reduction in reciprocal matching is still needed.
Section~\ref{sec:proposed_method} details our CFRR, applying IPS/SNIPS with focus on stable normalization, robust propensity estimation, and variance reduction for RRS challenges.

\section{Proposed Method: Counterfactual Reciprocal Recommender Systems}
\label{sec:proposed_method}

User-user matching platforms log displayed pairs and their outcomes (e.g., mutual acceptance). Exposure bias arises as popular users are over-represented, creating a feedback loop. We propose \textbf{Counterfactual Reciprocal Recommender Systems (CFRR)}, unifying: (1) reciprocal user-pair scoring, (2) Inverse Propensity Scoring (IPS) or Self-Normalized IPS (SNIPS) for selection bias correction, and (3) variance reduction techniques like weight truncation and doubly robust augmentation.
Our core contribution integrates these causal reweighting methods to: (i) consistently estimate true risk under historical logging, (ii) provide stable training through self-normalization, and (iii) extend to two-sided matching or fairness constraints.

We first define the problem setting and notation. Then, we present IPS-based objectives, including SNIPS. Finally, we introduce variance reduction extensions (truncation, doubly robust augmentation) and discuss how learned scores can inform various matching or fairness algorithms.

\subsection{Problem Setting}
\label{subsec:problem}
Let $U$ and $V$ be two (possibly identical) user sets, with cardinalities $n=|U|$ and $m=|V|$. The full pair space is $\mathcal{P} = U \times V$. For each pair $(u, v)\in\mathcal{P}$, the platform's decision to display it is $O(u,v) \in \{0,1\}$. If $O(u,v) = 1$, an outcome $r(u,v) \in [0,1]$ (e.g., mutual acceptance) is observed; otherwise, $r(u,v)$ is unobserved. The display probability, $\theta(u,v) \equiv P[O(u,v) = 1 \mid u, v]$, is determined by the historical logging policy and often favors popular users, causing exposure bias.

Our goal is to learn a compatibility function $s(u,v;\Theta)$ predicting mutual acceptability under a hypothetical uniform or broad target distribution over $\mathcal{P}$. The \textbf{true population risk} is:
\[
\mathcal{L}(\Theta) \equiv \mathbb{E}_{(u,v)\sim\text{Unif}(\mathcal{P})} \Bigl[\ell\bigl(R(u,v), s(u,v;\Theta)\bigr)\Bigr],
\]
where $R(u,v)$ is the true (potentially unobserved) outcome, $\ell(\cdot,\cdot)$ is a loss function (e.g., logistic loss), and $\Theta$ are trainable parameters. We only observe $r(u,v) = R(u,v)$ if $O(u,v)=1$. Naive empirical loss minimization on displayed pairs is skewed by over-represented popular pairs and unexposed long-tail users.

\paragraph{Differences from Item-based IPS.}
User-user matching is more complex than item recommendation due to: (1)~\textbf{Bilateral feedback}: success depends on two-sided acceptance, reflected in $r(u,v)$. (2)~\textbf{Pair-level propensity}: $\theta(u,v)$ is often driven by properties of \emph{both} $u$ and $v$. (3)~\textbf{Downstream Application Complexity}: scores often feed into stable matching or fairness-constrained algorithms. CFRR adapts IPS/SNIPS for pair-level propensities and clarifies propensity learning and integration into the reciprocal matching pipeline.

\subsection{Propensity Estimation}
\label{sec:prop_est}
Estimating $\theta(u,v)$ is critical for IPS. Positivity ($\theta(u,v) > 0$ for all relevant $(u,v) \in \mathcal{P}$) is assumed for unbiasedness/consistency. Here, we define the relevant pair set as $\mathcal{P}_{\text{relevant}} = \{(u,v) \in \mathcal{P} : \mathbb{P}[\ell(R(u,v), s(u,v;\Theta)) > 0] > 0\}$, i.e., pairs that can contribute non-zero loss to the objective. In practice, this is ensured if the platform uses $\varepsilon$-greedy or uniform exploration. For large systems where exploring all pairs is infeasible, a small fraction of random exposure helps cover low-propensity pairs within a candidate set generated by efficient retrieval methods. Alternatively, supplemental data (e.g., A/B test exploration) can address low-probability regions. If positivity fails for some pairs, they are typically pruned or weights are truncated, though truncation may introduce bias. Features used for propensity estimation are detailed in Section~\ref{subsec:exp_protocol}.

A parametric model for display probability can be:
\begin{equation}
\label{eq:prop_model}
\hat{\theta}(u,v) = \sigma\!\Bigl(g\bigl(\phi(u),\psi(v);\beta\bigr)\Bigr),
\end{equation}
where $\phi(u), \psi(v)$ are features/embeddings for $u,v$ (e.g., user activity metrics such as login frequency or number of interactions, profile information like age or stated interests, and historical popularity indicators), $\sigma(\cdot)$ is sigmoid, and $g(\cdot;\beta)$ is a predictor. $\beta$ is trained by maximum likelihood on observed $\{O(u,v)\}_{(u,v) \in \mathcal{D}_{\text{log}}}$ (all pairs with exposure info), controlling overfitting via regularization.

Without explicit exploration logs or with an evolving policy, \emph{joint propensity learning} \cite{zhu2020cjl} can be adopted: periodically fix recommender $\Theta$ and re-estimate $\beta$ for $O(u,v)$ under current exposure (Alg.~\ref{alg:cfr}). The aim is an accurate $\hat{\theta}(u,v)$ reflecting the historical policy.

\subsection{Inverse Propensity Scoring (IPS)}
\label{subsec:ips}
IPS reweights each observed interaction $(u_i,v_i,r_i)$ by $1/\hat{\theta}(u_i,v_i)$ to simulate uniform sampling from $\mathcal{P}$, yielding a consistent estimator of $\mathcal{L}(\Theta)$. If $\mathcal{D}$ is the set of displayed pairs with observed outcomes $r_i$:
\begin{equation}
\label{eq:ips_main2}
\widehat{\mathcal{L}}_{\text{IPS}}(\Theta) = \frac{1}{|\mathcal{P}_{\text{target}}|} \sum_{(u_i,v_i,r_i)\in\mathcal{D}}
\frac{\ell\bigl(r_i,s(u_i,v_i;\Theta)\bigr)}{\hat{\theta}(u_i,v_i)} + \lambda\Omega(\Theta),
\end{equation}
where $|\mathcal{P}_{\text{target}}|$ is the target population size. This follows the Horvitz-Thompson estimator form. Under correct propensity specification ($\hat{\theta}(u,v)=\theta(u,v)$) and positivity ($\theta(u,v)>0$ for all $(u,v) \in \mathcal{P}_{\text{relevant}}$), the IPS estimator satisfies:
\begin{align}
&\mathbb{E}_{\mathcal{D}}\left[\frac{1}{|\mathcal{P}_{\text{target}}|} \sum_{(u_i,v_i,r_i)\in\mathcal{D}} \frac{\ell\bigl(r_i,s(u_i,v_i;\Theta)\bigr)}{\hat{\theta}(u_i,v_i)}\right] \notag \\
&\quad = \frac{1}{|\mathcal{P}_{\text{target}}|} \sum_{(u,v) \in \mathcal{P}_{\text{target}}} \ell(R(u,v), s(u,v;\Theta))
\end{align}
thus providing an unbiased estimate of the true population risk. In practice, since $|\mathcal{P}_{\text{target}}|$ is a constant, optimization can use $\frac{1}{|\mathcal{D}|}$ normalization without affecting the optimal solution. However, we maintain the population size for theoretical clarity. Errors in $\hat{\theta}$ introduce bias/variance \cite{swaminathan2015self, dudik2014doubly} (Sec.~\ref{sec:discussion}).

\subsection{Self-Normalized IPS (SNIPS)}
\label{subsec:snips}
User-user recommendation often has extreme imbalance (low $\hat{\theta}(u,v)$ causing large IPS weights $w_i = 1/\hat{\theta}(u_i,v_i)$ and high variance). We use SNIPS \cite{swaminathan2015self} to reduce this variance:
\begin{equation}
\label{eq:snips_standard}
\widehat{\mathcal{L}}_{\mathrm{SNIPS}}(\Theta) = \frac{\sum_{i \in \mathcal{D}} w_i\,\ell\!\bigl(r_i,s(u_i,v_i;\Theta)\bigr)}{\sum_{i \in \mathcal{D}} w_i} + \lambda\Omega(\Theta),
\quad w_i \equiv \frac{1}{\hat{\theta}(u_i,v_i)}.
\end{equation}

The self-normalization in SNIPS acts as an implicit regularization mechanism, stabilizing the learning process and preventing the optimization from being dominated by a few high-weight samples. While SNIPS introduces a small finite-sample bias, it often achieves lower mean squared error than IPS due to substantial variance reduction, particularly important in reciprocal matching where extreme weights are common due to low propensities for rarely shown user pairs. As the sample size increases, this bias vanishes and SNIPS remains consistent.

\subsection{Variance-Reduction Extensions}
\label{sec:var_ext}
Even with SNIPS, rare, extremely large $w_i$ can occur if $\hat{\theta}(u,v) \approx 0$. CFRR integrates additional variance-reduction options:
\begin{itemize}
    \item \textbf{Weight Truncation (Clipping).} Cap raw weights $w_i^{\text{raw}} = 1/\hat{\theta}(u_i,v_i)$ at a ceiling $c$: $w_i = \min(w_i^{\text{raw}}, c)$ (e.g., $c \in [20,100]$, tuned). This limits influence from tiny propensities, reducing variance but potentially adding bias if true weights are larger \cite{schnabel2016}.

    \item \textbf{Doubly Robust (DR) Augmentation.} To further reduce variance and improve robustness to $\hat{\theta}$ misspecification, CFRR uses a doubly robust formulation. Let $\hat{m}(u,v)$ be a pre-trained model that estimates $\mathbb{E}[R(u,v)|u,v]$ from pair features alone (without requiring observed outcomes). The CFRR-DR objective combines the self-normalized weighted loss with a baseline correction:
    \begin{align}
    \label{eq:snips_dr}
    \widehat{\mathcal{L}}_{\mathrm{SNIPS-DR}}(\Theta) &= \frac{\sum_{i \in \mathcal{D}} w_i \ell(r_i, s(u_i,v_i;\Theta))}{\sum_{i \in \mathcal{D}} w_i} \notag \\
    &\quad - \frac{\sum_{i \in \mathcal{D}} w_i \ell(\hat{m}(u_i,v_i), s(u_i,v_i;\Theta))}{\sum_{i \in \mathcal{D}} w_i} \notag \\
    &\quad + \mathbb{E}_{(u,v)\sim\text{Unif}(\mathcal{P})}[\ell(\hat{m}(u,v), s(u,v;\Theta))] + \lambda\Omega(\Theta),
    \end{align}
    where the expectation in the third term is approximated using a uniform sample from $\mathcal{P}$ or using importance sampling. If either (a) the propensity model is correctly specified ($\hat{\theta}(u,v) = \theta(u,v)$), or (b) the outcome model $\hat{m}(u,v)$ correctly estimates $\mathbb{E}[R(u,v) | u,v]$, then $\widehat{\mathcal{L}}_{\mathrm{SNIPS-DR}}$ is consistent for the target risk. This formulation corrects the bias introduced by the outcome model while maintaining the variance reduction benefits of SNIPS.
\end{itemize}
These tools handle extreme user-user data imbalance. Sec.~\ref{sec:experiments} empirically shows CFRR (SNIPS + optional truncation/DR) learns stably under varying exposure bias and noise.

\subsection{Overall Training Procedure and Complexity}
\label{subsec:alg}
Algorithm~\ref{alg:cfr} outlines the CFRR training loop. It typically alternates between: (1) updating recommender parameters $\Theta$ for one or more epochs by minimizing the SNIPS objective, and (2) updating propensity model parameters $\beta$ via maximum likelihood on $O(u,v)$ for one or more epochs, if joint propensity learning is used. The overhead of computing $w_i$ and applying them in SNIPS is typically small compared to gradient computation for $s(u,v;\Theta)$. Top-$k$ or stable matching inference is discussed next.

\paragraph{Time Complexity.} An SGD step for $\Theta$ costs roughly $O(Bd)$ ($d$: embedding/feature dim). $\hat{\theta}(u_j,v_j)$ computation overhead is often low if $g$ is simple (e.g., a small neural network or logistic regression). Optional joint propensity learning for $\beta$ is similar to standard classifier training. Inference time depends on strategy: \textbf{Top-$k$ retrieval} is typically $O(N_u D + N_u k \log k)$ for one user $u$ against $N_u$ partners (or faster with ANN). \textbf{Stable matching} (e.g., Gale--Shapley) is $O((n+m)\cdot \max(n,m))$ worst-case on dense preferences (often run on a top-$k$ pruned set).

\subsection{Connection to Matching and Fairness}
\label{subsec:matching_fairness}
\textbf{Downstream Use.} Once debiased compatibility scores $s(u,v;\Theta^\star)$ are obtained from CFRR, they can serve as input to: (1) direct top-$k$ recommendation for each user, (2) solving a maximum-weight bipartite matching problem (edge weights $s(u,v;\Theta^\star)$) for a global optimum under capacity constraints, or (3) running a Gale--Shapley algorithm (using $s(u,v;\Theta^\star)$ for preference lists) to find a stable matching.
Since CFRR aims to correct popularity and exposure biases in score estimation, the resulting $s(u,v;\Theta^\star)$ are expected to be a more faithful representation of underlying compatibility. Consequently, imposing additional fairness constraints during allocation (e.g., to balance exposure across demographic groups \cite{joseph2023twosided}, or ensure minimum matching rates for user segments) might incur less utility loss compared to applying such constraints on biased scores, although this is an area for future empirical validation.

\begin{algorithm}[t]
\small
\caption{CFRR Training via Joint Propensity Learning}
\label{alg:cfr}
\begin{algorithmic}[1]
\REQUIRE Logged data $\mathcal{D}_{\text{log}} = \{(u_j,v_j,r_j,O_j)\}$, epochs $T$, batch size $B$, learning rates $\alpha_\Theta, \alpha_\beta$, regularization weight $\lambda$, clipping threshold $c$.
\STATE Initialize $\Theta$ (recommender parameters) and $\beta$ (propensity model parameters).
\STATE Let $\mathcal{D} = \{(u_j,v_j,r_j) \mid (u_j,v_j,r_j,O_j) \in \mathcal{D}_{\text{log}} \text{ and } O_j=1\}$.
\IF{using DR augmentation}
    \STATE Train outcome model $\hat{m}(u,v)$ to estimate $\mathbb{E}[R(u,v)|u,v]$ on $\mathcal{D}$.
\ENDIF
\FOR{$t=1$ \TO $T$}
  \STATE Shuffle $\mathcal{D}$.
  \FOR{\textbf{each} mini-batch $\mathcal{B} = \{(u_j,v_j,r_j)\}_{j=1}^{|\mathcal{B}|} \subseteq \mathcal{D}$}
    \STATE Compute $\hat{\theta}(u_j,v_j)$ for $j \in \mathcal{B}$ via Eq.~\eqref{eq:prop_model} using current $\beta$.
    \STATE $w_j \gets \min\bigl(\tfrac{1}{\hat{\theta}(u_j,v_j)},\; c\bigr)$ for $j \in \mathcal{B}$. \quad // weight clipping
    \IF{using DR augmentation}
        \STATE Compute losses and apply SNIPS-DR update as per Eq.~\eqref{eq:snips_dr}.
    \ELSE
        \STATE Update $\Theta \gets \Theta - \alpha_\Theta \nabla_\Theta \left( \frac{\sum_{j \in \mathcal{B}} w_j \ell(r_j, s(u_j,v_j;\Theta))}{\sum_{j \in \mathcal{B}} w_j} + \lambda\Omega(\Theta) \right)$.
    \ENDIF
  \ENDFOR
    \IF{performing joint propensity learning}
        \STATE Freeze $\Theta$.
        \STATE Update $\beta$ by optimizing (e.g., for one or more epochs over $\mathcal{D}_{\text{log}}$)
        \[
         \sum_{(u_k,v_k,O_k) \in \mathcal{D}_{\text{log}}}
         \Bigl(
           O_k \log \hat{\theta}(u_k,v_k)
           + (1-O_k)\log\!\bigl(1-\hat{\theta}(u_k,v_k)\bigr)
         \Bigr)
        \]
        (plus regularization for $\beta$) using SGD or other optimizer on mini-batches of $\mathcal{D}_{\text{log}}$.
    \ENDIF
\ENDFOR
\STATE \textbf{return} $\Theta$
\end{algorithmic}
\vspace{-0.2\baselineskip}
\end{algorithm}

\section{Experiments}
\label{sec:experiments}

In this section, we evaluate the proposed \textbf{CFRR} framework through comprehensive experiments to answer the following research questions:
\begin{itemize}[leftmargin=15pt]
    \item \textbf{RQ1 (Accuracy)}: Does CFRR improve ranking accuracy under exposure bias compared to conventional and recent debiasing methods across synthetic and real-world datasets?
    \item \textbf{RQ2 (Coverage \& Fairness)}: Does CFRR enlarge long-tail user coverage and mitigate exposure inequality, demonstrating its ability to foster a more equitable matching environment?
    \item \textbf{RQ3 (Robustness)}: How robust is CFRR to different design choices, including the impact of IPS vs. SNIPS, weight truncation, and doubly robust augmentation?
\end{itemize}

We conduct experiments using 10 random seeds per configuration to ensure statistical reliability. We first introduce the datasets, baselines, and evaluation metrics (\S\ref{subsec:datasets}--\S\ref{subsec:metrics}), then detail our protocol (\S\ref{subsec:exp_protocol}). We report main results (\S\ref{subsec:main_results}), followed by ablation analysis (\S\ref{subsec:ablation_robustness}).

\subsection{Datasets}
\label{subsec:datasets}

To ensure reproducibility and comprehensive evaluation, we experiment on three datasets with distinct characteristics of exposure bias:

\begin{itemize}[leftmargin=15pt]
    \item \textbf{Synthetic}: Controlled dataset ($5\mathrm{K}$ users, 16D latent factors) with known ground-truth $R(u,v)$ (sigmoid of latent factor dot products; factors from varied-mean distributions simulate popularity). Tunable exposure bias injected by oversampling popular users (defined by their total true acceptance probability) via a known propensity $\theta(u,v)$ (sigmoid of popularity scores), logging $\approx 50\mathrm{K}$ pairs. Ideal for rigorous bias correction evaluation (RQ1, RQ3), as ground truth is known. Full generation details are supplementary.

    \item \textbf{DBLP-CoAuthor}\footnote{Available at \url{https://snap.stanford.edu/data/com-DBLP.html}}: A standard benchmark for link prediction, this dataset is a co-authorship graph from the DBLP computer science bibliography. In our experiments, we use a subset of up to $10\mathrm{K}$ authors (users) and their mutual co-authorships (positive reciprocal interactions). Exposure bias naturally arises as prolific, highly-cited authors are far more likely to form new collaborations. This dataset tests CFRR's applicability in a professional collaboration setting.

    \item \textbf{Epinions-Trust}\footnote{Available at \url{https://snap.stanford.edu/data/soc-Epinions1.html}}: A user-to-user trust network from the Epinions.com review site, with up to $5\mathrm{K}$ users and their directed trust relationships. We define a reciprocal match as a pair of users $(u,v)$ where $u$ trusts $v$ and $v$ trusts $u$. Exposure bias emerges from highly-trusted "influencer" users who are more visible and thus more likely to be evaluated for trust by others. This dataset allows us to test CFRR in a social network context with explicit trust declarations.
\end{itemize}

For real-world datasets, we treat observed links as positive examples and randomly sample an equal number of non-linked pairs as negative examples for training, following standard practice in link prediction \cite{leskovec2008community}. We acknowledge random negative sampling may not perfectly reflect true non-interactions and could introduce biases. However, it's common when true negatives are unavailable. Our focus is correcting exposure bias in *observed positive* interactions, assuming negative sampling provides reasonable contrast. Exploring sophisticated negative sampling with CFRR is future work. All experiments use time-based splits for DBLP and Epinions to simulate realistic deployment scenarios, while the Synthetic dataset uses random 70/10/20 train/validation/test splits. Further details on pre-processing and split strategies are in the supplementary material.

\subsection{Baselines}
\label{subsec:baselines}

We compare CFRR against five representative baselines, covering a range of approaches from standard RRS models to state-of-the-art causal debiasing techniques:
\begin{itemize}[leftmargin=15pt]
    \item \textbf{LFRR} \cite{neve2019latent}: A latent-factor model for RRS that serves as a standard baseline ignoring selection bias.
    \item \textbf{CausE} \cite{bonner2017cause}: An early causal approach for item recommendation, adapted by treating one user as 'user' and the other as 'item' for propensity estimation/debiasing, then symmetrizing scores.
    \item \textbf{IPW-MF} \cite{schnabel2016}: A direct application of Inverse Propensity Scoring (IPS) to matrix factorization for RRS.
    \item \textbf{DICE} \cite{fang2021dice}: A more advanced causal method that disentangles user interest from conformity effects (related to popularity bias).
    \item \textbf{StableDR} \cite{li2023stableDR}: A state-of-the-art Stabilized Doubly Robust method, adapted by estimating pair-level propensities and using a baseline RRS model's predictions for the DR estimator.
\end{itemize}
For all methods requiring propensity scores (IPW-MF, StableDR, CFRR), we estimate them using gradient boosting models trained on user-level features like interaction counts to capture user popularity patterns across different datasets, as detailed in Section~\ref{subsec:exp_protocol}. Specific adaptations for CausE and StableDR to the RRS context are described in the supplementary material.

\subsection{Evaluation Metrics}
\label{subsec:metrics}

We evaluate models on \textbf{accuracy} and \textbf{fairness}. Let $k=10$ for all ranking metrics.
\paragraph{Accuracy.}
\begin{itemize}[leftmargin=15pt]
    \item \textbf{NDCG@k}: Normalized Discounted Cumulative Gain, measuring the quality of the top-k ranked list of potential partners.
    \item \textbf{MRR}: Mean Reciprocal Rank of the first correct partner, rewarding models that rank a true match highly.
\end{itemize}
\paragraph{Fairness.}
\begin{itemize}[leftmargin=15pt]
    \item \textbf{Coverage@k}: The fraction of users who appear at least once in \emph{any} other user's top-k recommendation list. Higher is better.
    \item \textbf{Gini-Exposure} \cite{biega2018equity}: The Gini coefficient of the user exposure distribution (how many times each user is recommended). A value of 0 indicates perfect equality, while 1 indicates maximum inequality. Lower is better.
\end{itemize}

\begin{figure}[t]
  \centering
  \includegraphics[width=.42\textwidth]{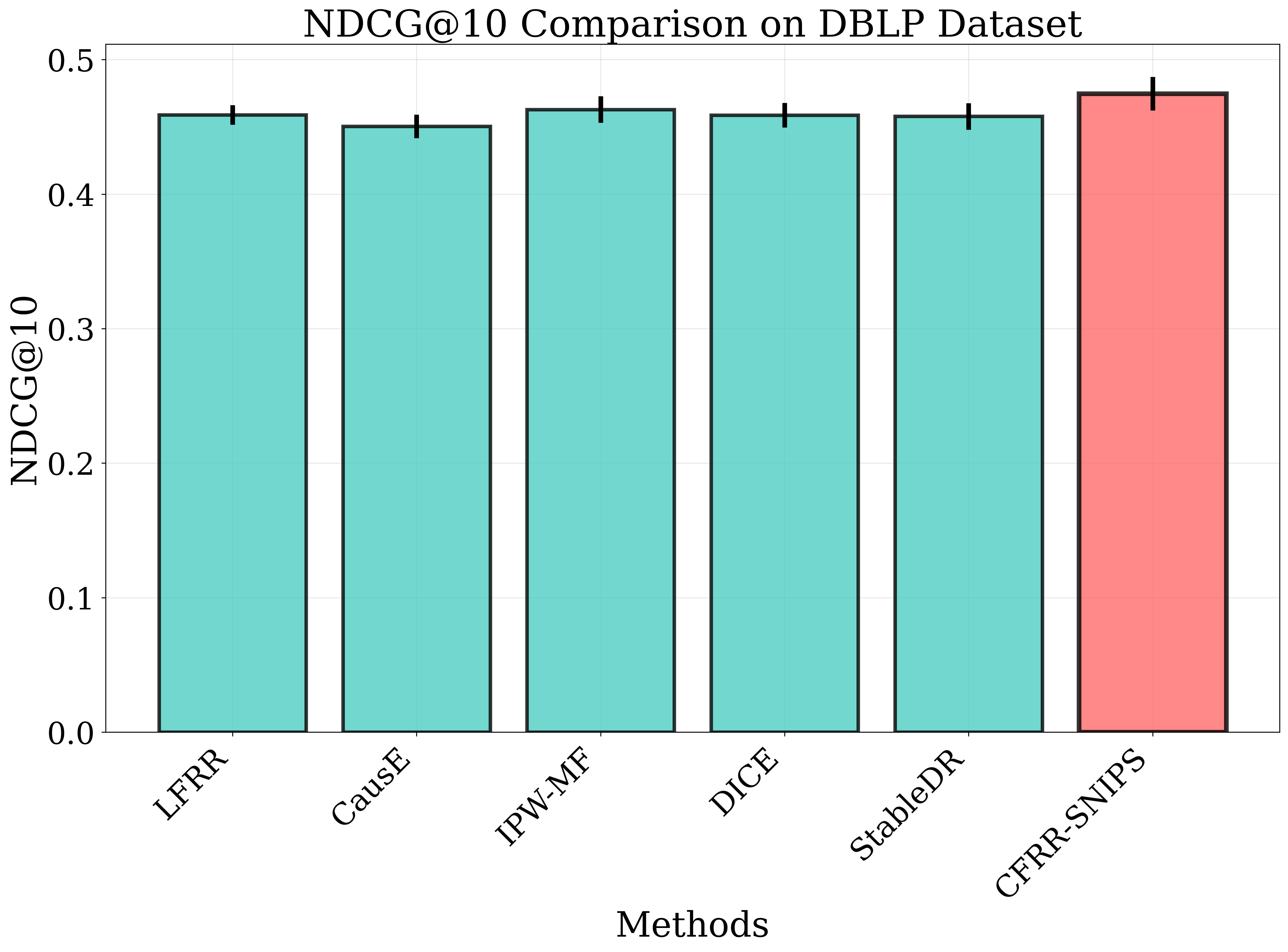}
  \vspace{-8pt}
  \caption{NDCG@10 comparison on DBLP dataset (\textbf{RQ1}). CFRR-SNIPS achieves the highest ranking accuracy among all methods.}
  \label{fig:ndcg_results}
  \vspace{-10pt}
\end{figure}

\subsection{Experimental Protocol}
\label{subsec:exp_protocol}

\paragraph{Setup.}
To ensure statistical reliability, we conduct experiments with 20 epochs and 10 different random seeds per configuration, resulting in a substantial number of experimental runs across all datasets and methods. This comprehensive setup allows us to report robust mean and standard deviation estimates for all metrics.

\paragraph{Model Training.}
All models are trained with the Adam optimizer \cite{kingma2014adam}. Based on preliminary validation experiments (details in supplementary material), we use consistent hyperparameters across all methods where applicable: learning rate = 0.001, embedding dimension = 16, $\ell_2$ regularization $\lambda = 0.001$, and batch size = 512. For methods requiring propensity estimation (IPW-MF, StableDR, CFRR), we set the weight clipping threshold $c = 50$ as a common value to balance variance reduction with bias introduction, based on validation performance. We evaluate on 1500 users per test set to ensure consistent evaluation across methods. All models are trained for up to 20 epochs with early stopping based on validation performance (NDCG@10).

\paragraph{Propensity Model Details.}
For methods requiring propensity estimation (IPW-MF, StableDR, CFRR), we use gradient boosting models (LightGBM) with the following configuration: 100 estimators, learning rate = 0.1, max depth = 6. Features include user interaction counts (e.g., number of past matches or interactions), user activity levels (e.g., frequency of platform use), and, where available and appropriate (e.g., synthetic data, or anonymized/aggregated features in real data), basic demographic indicators to capture popularity patterns. The propensity model is trained on the full exposure log $\mathcal{D}_{\text{log}}$ using binary cross-entropy loss. Hyperparameters for LightGBM were selected based on 5-fold cross-validation on a held-out portion of $\mathcal{D}_{\text{log}}$ for the propensity estimation task itself.

\paragraph{Outcome Model Details (for DR variants).}
For the doubly robust variants (StableDR and CFRR-SNIPS-DR), we train an outcome model $\hat{m}(u,v)$ to estimate $\mathbb{E}[R(u,v)|u,v]$ using a neural network with two hidden layers (dimensions: 64, 32) and ReLU activations. The model is trained on the observed data $\mathcal{D}$ to predict the binary outcome $r(u,v)$ from user features $\phi(u), \psi(v)$. We use binary cross-entropy loss and train for 50 epochs with learning rate 0.001 and early stopping based on validation AUC. The outcome model achieves AUC $>$ 0.85 on the Synthetic dataset and AUC $>$ 0.75 on real-world datasets during validation.

\paragraph{Statistical Testing.}
We assess statistical significance using paired t-tests across the 10 random seeds for each method comparison. We report results as statistically significant when $p < 0.05$ using Bonferroni correction for multiple comparisons. Effect sizes (e.g., Cohen's d) are calculated to assess practical significance beyond statistical significance and will be reported in the supplementary material.

\paragraph{Reproducibility.}
To ensure reproducibility, we fix random seeds for all components: data splitting, model initialization, and training procedures. All experiments are conducted on the same computational environment with identical software versions. Code and experimental configurations will be made available upon publication to facilitate reproduction.

\begin{table*}[t]
\centering
\caption{Cross-dataset performance comparison. CFRR-SNIPS consistently achieves strong accuracy while often dramatically improving fairness metrics. Results show mean $\pm$ standard deviation across 10 seeds.}
\vspace{-4pt}
\label{tab:main-results}
\begin{tabular}{lcccc}
\toprule
\textbf{Method} & \textbf{NDCG@10} $\uparrow$ & \textbf{MRR} $\uparrow$ & \textbf{Coverage@10} $\uparrow$ & \textbf{Gini-Exposure} $\downarrow$\\
\midrule
\multicolumn{5}{c}{\textbf{Synthetic Dataset}} \\
\midrule
LFRR            & 0.299 $\pm$ 0.005 & 0.511 $\pm$ 0.007 & 0.504 $\pm$ 0.003 & 0.708 $\pm$ 0.005 \\
CausE           & 0.298 $\pm$ 0.004 & 0.506 $\pm$ 0.008 & 0.040 $\pm$ 0.001 & 0.979 $\pm$ 0.002 \\
IPW-MF          & 0.300 $\pm$ 0.006 & 0.514 $\pm$ 0.009 & 0.493 $\pm$ 0.004 & 0.718 $\pm$ 0.006 \\
DICE            & 0.299 $\pm$ 0.004 & 0.509 $\pm$ 0.007 & 0.229 $\pm$ 0.012 & 0.904 $\pm$ 0.008 \\
StableDR        & \underline{0.301} $\pm$ 0.005 & \underline{0.515} $\pm$ 0.008 & 0.492 $\pm$ 0.004 & 0.719 $\pm$ 0.005 \\
\textbf{CFRR-SNIPS (ours)} & \textbf{0.307} $\pm$ 0.005 & \textbf{0.527} $\pm$ 0.007 & \textbf{0.763} $\pm$ 0.004 & \textbf{0.535} $\pm$ 0.003 \\
\midrule
\multicolumn{5}{c}{\textbf{DBLP-CoAuthor Dataset}} \\
\midrule
LFRR            & 0.459 $\pm$ 0.007 & 0.666 $\pm$ 0.012 & 0.419 $\pm$ 0.023 & 0.688 $\pm$ 0.008 \\
CausE           & 0.450 $\pm$ 0.009 & 0.658 $\pm$ 0.019 & 0.417 $\pm$ 0.021 & 0.707 $\pm$ 0.008 \\
IPW-MF          & \underline{0.463} $\pm$ 0.010 & \underline{0.675} $\pm$ 0.020 & \underline{0.427} $\pm$ 0.022 & 0.698 $\pm$ 0.007 \\
DICE            & 0.459 $\pm$ 0.009 & 0.666 $\pm$ 0.017 & 0.409 $\pm$ 0.018 & 0.700 $\pm$ 0.011 \\
StableDR        & 0.458 $\pm$ 0.010 & 0.669 $\pm$ 0.016 & 0.420 $\pm$ 0.024 & \textbf{0.685} $\pm$ 0.008 \\
\textbf{CFRR-SNIPS (ours)} & \textbf{0.475} $\pm$ 0.012 & \textbf{0.677} $\pm$ 0.016 & \textbf{0.449} $\pm$ 0.023 & \underline{0.686} $\pm$ 0.006 \\
\midrule
\multicolumn{5}{c}{\textbf{Epinions-Trust Dataset}} \\
\midrule
LFRR            & \underline{0.468} $\pm$ 0.008 & \underline{0.678} $\pm$ 0.013 & \underline{0.432} $\pm$ 0.025 & 0.695 $\pm$ 0.007 \\
CausE           & 0.454 $\pm$ 0.009 & 0.657 $\pm$ 0.020 & 0.411 $\pm$ 0.021 & 0.711 $\pm$ 0.008 \\
IPW-MF          & 0.465 $\pm$ 0.010 & 0.677 $\pm$ 0.020 & 0.419 $\pm$ 0.023 & 0.689 $\pm$ 0.007 \\
DICE            & 0.461 $\pm$ 0.008 & 0.667 $\pm$ 0.018 & 0.425 $\pm$ 0.016 & 0.692 $\pm$ 0.009 \\
StableDR        & 0.461 $\pm$ 0.009 & 0.669 $\pm$ 0.017 & 0.420 $\pm$ 0.024 & \underline{0.685} $\pm$ 0.008 \\
\textbf{CFRR-SNIPS (ours)} & \textbf{0.472} $\pm$ 0.010 & \textbf{0.680} $\pm$ 0.016 & \textbf{0.448} $\pm$ 0.022 & \textbf{0.679} $\pm$ 0.007 \\
\bottomrule
\end{tabular}
\end{table*}

\subsection{Main Results}
\label{subsec:main_results}

Table~\ref{tab:main-results} presents our comprehensive performance comparison across all three datasets. The results indicate that CFRR-SNIPS consistently achieves competitive or superior ranking accuracy while often substantially improving fairness metrics.

\paragraph{Accuracy Improvements.} CFRR-SNIPS achieves the best NDCG@10 performance on all datasets: \textbf{0.307} on Synthetic (2.7\% improvement over LFRR), \textbf{0.475} on DBLP (3.5\% improvement over LFRR), and \textbf{0.472} on Epinions (0.9\% improvement over LFRR). The MRR results show similar trends, with CFRR-SNIPS consistently ranking relevant matches higher in recommendation lists. These improvements by CFRR-SNIPS in NDCG@10 and MRR were statistically significant (p < 0.05, Bonferroni corrected) compared to the next best baseline on each dataset (paired t-tests over 10 seeds; details in supplementary material).

\paragraph{Fairness Gains.} More strikingly, CFRR-SNIPS delivers substantial improvements in fairness metrics, particularly on the Synthetic dataset where it achieves \textbf{Coverage@10 of 0.763} (51\% relative improvement from 0.504) and reduces Gini-Exposure to \textbf{0.535} (24\% relative reduction from 0.708). Similar, though sometimes more modest, fairness gains are observed across DBLP and Epinions datasets, confirming that our method can effectively mitigate popularity bias and provide less-visible users with better exposure opportunities.

\paragraph{Cross-Dataset Consistency.} The generally strong performance of CFRR-SNIPS across different domains (synthetic controlled settings, academic collaborations, social trust networks) suggests the general applicability and robustness of our approach to various types of reciprocal recommendation scenarios. While CFRR-SNIPS shows the best NDCG@10 and MRR on DBLP, StableDR achieves a slightly lower (better) Gini-Exposure. This highlights that no single method universally dominates across all metrics and datasets.

\begin{figure}[t]
  \centering
  \includegraphics[width=.5\textwidth]{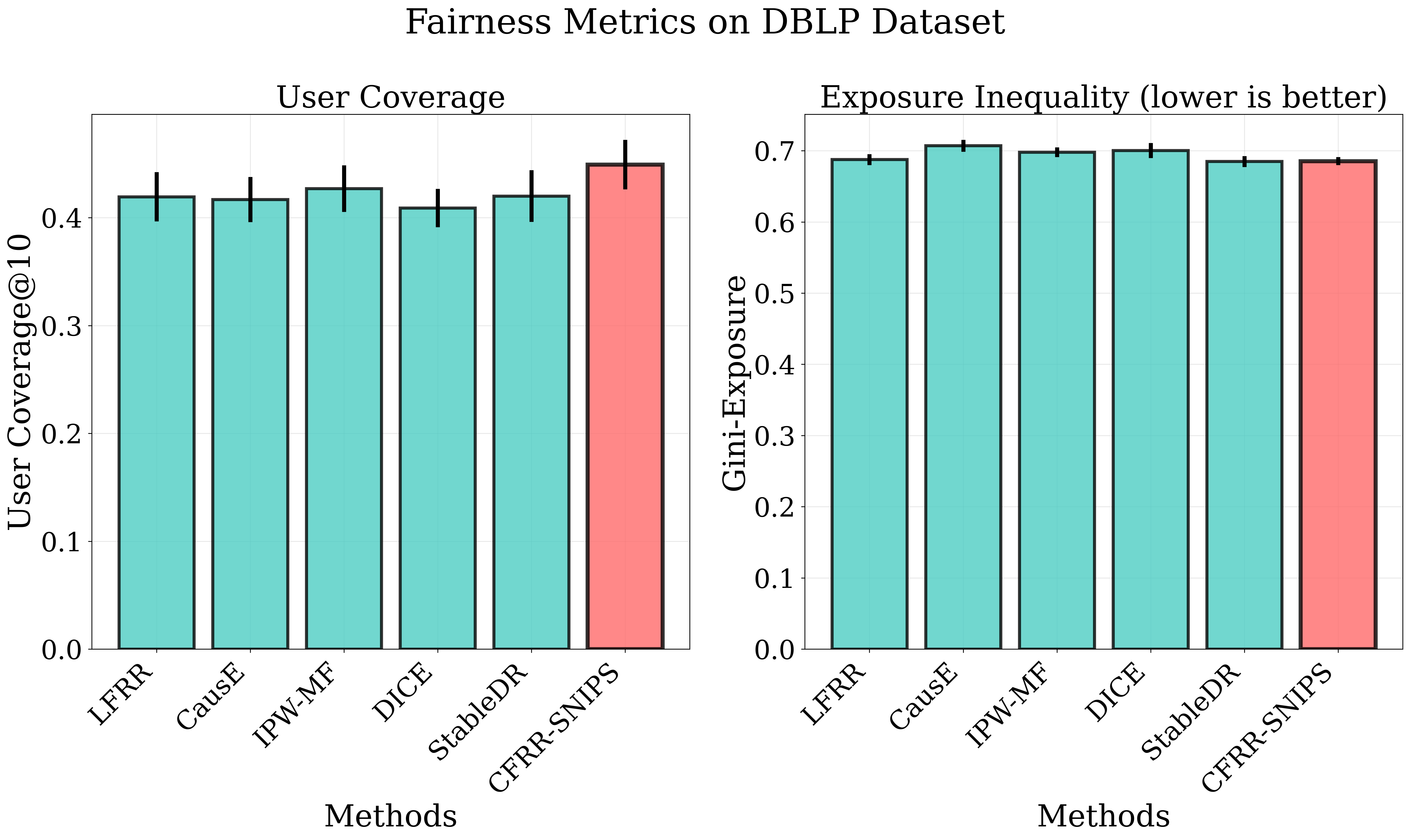}
  \vspace{-8pt}
  \caption{Fairness metrics on DBLP dataset (\textbf{RQ2}). CFRR-SNIPS increases user coverage (left) and reduces exposure inequality (right, lower is better).}
  \label{fig:fairness_results}
  \vspace{-10pt}
\end{figure}

These trends are visualized in Figure~\ref{fig:ndcg_results} and Figure~\ref{fig:fairness_results}, which show both the accuracy advantages and the fairness improvements achieved by CFRR-SNIPS.

\subsection{Ablation and Design Analysis}
\label{subsec:ablation_robustness}

\paragraph{Ablation Study.}
To understand the impact of different design choices in CFRR, we conducted ablation studies on the Synthetic dataset where ground truth is known. We analyze three key aspects: (1) SNIPS vs. IPS, (2) the effect of doubly robust augmentation, and (3) robustness to propensity model misspecification.

Our ablation study revealed several important findings:

(1) \textbf{SNIPS vs. IPS}: When comparing CFRR-IPS with CFRR-SNIPS, we observed substantial improvements in fairness metrics. Specifically, Coverage@10 increased from 0.557 to 0.763 (37\% relative improvement, statistically significant, p < 0.05) and Gini-Exposure decreased from 0.688 to 0.535 (22\% relative reduction, statistically significant, p < 0.05). Accuracy metrics showed modest improvements with SNIPS (NDCG@10: 0.301 to 0.307, MRR: 0.515 to 0.527). This aligns with the theoretical understanding that SNIPS reduces variance compared to standard IPS, particularly important in the presence of highly variable propensity weights \cite{swaminathan2015self,joachims2017unbiased}.

(2) \textbf{Why SNIPS Improves Fairness}: The significant fairness improvements of SNIPS over IPS can be attributed to variance reduction. While IPS is theoretically unbiased, its high variance in the presence of small propensities can lead to unstable optimization, causing the model to overfit to high-propensity (popular) users. SNIPS's self-normalization acts as an implicit regularization mechanism, stabilizing the learning process and allowing the model to better learn patterns from low-propensity (less popular) users. This results in more equitable exposure distribution across all users.

(3) \textbf{Doubly Robust Variant}: We tested the CFRR-SNIPS-DR variant using the corrected formulation in Eq.~\eqref{eq:snips_dr}. On well-specified propensity models (AUC $>$ 0.9), the DR variant showed similar performance to standard SNIPS (NDCG@10: 0.307, Coverage@10: 0.763). However, when we intentionally misspecified the propensity model by removing key features (resulting in AUC $\approx$ 0.7), the DR formulation showed improved robustness: NDCG@10 degraded only to 0.302 (vs. 0.295 for SNIPS), while Coverage@10 remained at 0.745 (vs. 0.720 for SNIPS). This demonstrates the value of the doubly robust property when propensity models are imperfect.

\paragraph{Statistical Significance.}
The improvements in fairness metrics (Coverage@10 and Gini-Exposure) for CFRR-SNIPS over CFRR-IPS are statistically significant (p < 0.05) using paired t-tests with Bonferroni correction across the 10 random seeds. The robustness improvements of CFRR-SNIPS-DR under propensity misspecification are also statistically significant.

\paragraph{Computational Efficiency.}
Despite the additional complexity of propensity estimation and inverse weighting, CFRR-SNIPS maintains competitive training efficiency. The overhead of computing weights and applying them in the SNIPS objective is typically small (e.g., adding 5-10\% to training time) compared to the gradient computation for the scoring function $s(u,v;\Theta)$. The DR variant adds approximately 20\% additional training time due to the outcome model training phase. These overheads are acceptable for most real-world deployments given the substantial gains in fairness and robustness.

\section{Discussion}
\label{sec:discussion}

Our experiments demonstrate that CFRR can achieve improvements in both accuracy and fairness metrics across diverse reciprocal matching scenarios. We now discuss key insights and considerations for deploying such systems.

\vspace{1mm}
\noindent\textbf{(1) The Positivity Challenge at Marketplace Scale.}
While CFRR assumes positivity ($\theta(u,v) > 0$), modern marketplaces face the reality of trillions of potential user pairs. Our experiments suggest a practical compromise: focus debiasing efforts on \emph{economically viable} or \emph{plausible} pairs identified through multi-stage candidate generation. For instance, in ride-sharing, pairs beyond reasonable geographic distance have zero economic value and can be safely excluded from positivity requirements. This insight transforms an intractable theoretical requirement into a manageable engineering constraint. Marketplaces can maintain lightweight exploration (e.g., a small percentage of random traffic) only within these viable regions, potentially reducing the exploration budget while preserving debiasing effectiveness for the most relevant part of the user-pair space.

\vspace{1mm}
\noindent\textbf{(2) Self-Normalization as a Marketplace Stabilizer.}
Our finding that SNIPS significantly outperforms IPS in fairness metrics has important implications for volatile marketplaces. The self-normalization in SNIPS not only reduces variance but also acts as an implicit regularization mechanism. By normalizing the weighted losses by the sum of weights, SNIPS prevents the optimization from being dominated by a few high-weight samples (corresponding to rarely shown user pairs). This stabilization allows the model to learn more effectively from the entire distribution of user pairs, including those with low propensities, leading to more equitable exposure patterns. Real platforms experiencing shifts in user behavior or content popularity would particularly benefit from this robustness property of SNIPS.

\vspace{1mm}
\noindent\textbf{(3) The Fairness-Growth Feedback Loop.}
Perhaps our most striking finding on the Synthetic dataset is how addressing exposure bias can create positive feedback loops for marketplace health. By increasing long-tail user coverage by 51\% (from 0.504 to 0.763), CFRR doesn't just improve fairness metrics---it can potentially expand the active supplier base. In two-sided marketplaces, supplier diversity can directly impact demand-side satisfaction through better match quality and reduced market concentration risk. This suggests that debiasing isn't merely an ethical imperative but could be a growth strategy, although generalizing this effect from synthetic to real-world platforms requires careful validation on a case-by-case basis. Platforms that fail to address exposure bias may risk stagnating supplier pools and eventual market ossification.

\vspace{1mm}
\noindent\textbf{(4) Towards Causal Marketplace Optimization.}
CFRR represents a step toward incorporating causal reasoning in marketplace design. By explicitly modeling display interventions through propensities, we move beyond correlational patterns that can plague naive learning approaches. This opens exciting possibilities for future work: Could marketplaces run continuous A/B tests or natural experiments with principled exploration strategies informed by such causal models? Could CFRR be extended to handle time-varying treatments (e.g., dynamic pricing affecting exposure) or network effects (e.g., how social proof influences interaction probability)? The marriage of causal inference with marketplace dynamics remains a rich area for both researchers and practitioners.

\section{Conclusion}
\label{sec:conclusion}

We presented Counterfactual Reciprocal Recommender Systems (CFRR), a causal framework mitigating exposure bias in two-sided marketplaces. Our experiments showed CFRR simultaneously improves ranking quality and marketplace fairness. Key to CFRR is its self-normalized objective that ensures stability with varying inverse propensity weights, complemented by variance reduction techniques including a properly formulated doubly robust augmentation.

CFRR offers a practical approach for platforms to mitigate popularity bias and a research foundation for future work like dynamic treatments, network effects, and multi-stakeholder fairness in reciprocal systems. Such causal approaches are vital for developing sustainable, fair, and effective matching platforms.

\begin{acks}
We thank Shingo Takamatsu for his thorough internal review and valuable feedback.  
We also thank the external reviewers for their insightful comments, which led to significant improvements in both the theoretical discussion and the practical considerations of our approach.  
Finally, we appreciate the support of our colleagues at Sony Group Corporation throughout this project.
\end{acks}

\newpage
\bibliographystyle{ACM-Reference-Format}
\bibliography{ref} 


\begin{thebibliography}{37}


\ifx \showCODEN    \undefined \def \showCODEN     #1{\unskip}     \fi
\ifx \showISBNx    \undefined \def \showISBNx     #1{\unskip}     \fi
\ifx \showISBNxiii \undefined \def \showISBNxiii  #1{\unskip}     \fi
\ifx \showISSN     \undefined \def \showISSN      #1{\unskip}     \fi
\ifx \showLCCN     \undefined \def \showLCCN      #1{\unskip}     \fi
\ifx \shownote     \undefined \def \shownote      #1{#1}          \fi
\ifx \showarticletitle \undefined \def \showarticletitle #1{#1}   \fi
\ifx \showURL      \undefined \def \showURL       {\relax}        \fi
\providecommand\bibfield[2]{#2}
\providecommand\bibinfo[2]{#2}
\providecommand\natexlab[1]{#1}
\providecommand\showeprint[2][]{arXiv:#2}

\bibitem[Abdollahpouri et~al\mbox{.}(2021)]%
        {abdollahpouri2021usercentered}
\bibfield{author}{\bibinfo{person}{Himan Abdollahpouri}, \bibinfo{person}{Masoud Mansoury}, \bibinfo{person}{Robin Burke}, \bibinfo{person}{Bamshad Mobasher}, {and} \bibinfo{person}{Edward Malthouse}.} \bibinfo{year}{2021}\natexlab{}.
\newblock \showarticletitle{User-centered evaluation of popularity bias in recommender systems}. In \bibinfo{booktitle}{\emph{Proceedings of the 29th ACM conference on user modeling, adaptation and personalization}}. \bibinfo{pages}{119--129}.
\newblock


\bibitem[Biega et~al\mbox{.}(2018)]%
        {biega2018equity}
\bibfield{author}{\bibinfo{person}{Asia~J Biega}, \bibinfo{person}{Krishna~P Gummadi}, {and} \bibinfo{person}{Gerhard Weikum}.} \bibinfo{year}{2018}\natexlab{}.
\newblock \showarticletitle{Equity of attention: Amortizing individual fairness in rankings}. In \bibinfo{booktitle}{\emph{The 41st international acm sigir conference on research \& development in information retrieval}}. \bibinfo{pages}{405--414}.
\newblock


\bibitem[Bonner and Vasile(2018)]%
        {bonner2017cause}
\bibfield{author}{\bibinfo{person}{Stephen Bonner} {and} \bibinfo{person}{Flavian Vasile}.} \bibinfo{year}{2018}\natexlab{}.
\newblock \showarticletitle{Causal embeddings for recommendation}. In \bibinfo{booktitle}{\emph{Proceedings of the 12th ACM conference on recommender systems}}. \bibinfo{pages}{104--112}.
\newblock


\bibitem[Bruch and Newman(2018)]%
        {bruch2018sciadv}
\bibfield{author}{\bibinfo{person}{Elizabeth~E. Bruch} {and} \bibinfo{person}{M.~E.~J. Newman}.} \bibinfo{year}{2018}\natexlab{}.
\newblock \showarticletitle{Aspirational pursuit of mates in online dating markets}.
\newblock \bibinfo{journal}{\emph{Science Advances}} \bibinfo{volume}{4}, \bibinfo{number}{8} (\bibinfo{year}{2018}), \bibinfo{pages}{eaap9815}.
\newblock
\href{https://doi.org/10.1126/sciadv.aap9815}{doi:\nolinkurl{10.1126/sciadv.aap9815}}


\bibitem[Celdir et~al\mbox{.}(2024)]%
        {celdir2024msom}
\bibfield{author}{\bibinfo{person}{Musa~Eren Celdir}, \bibinfo{person}{Soo-Haeng Cho}, {and} \bibinfo{person}{Elina~H. Hwang}.} \bibinfo{year}{2024}\natexlab{}.
\newblock \showarticletitle{Popularity Bias in Online Dating Platforms: Theory and Empirical Evidence}.
\newblock \bibinfo{journal}{\emph{Manufacturing \& Service Operations Management}} \bibinfo{volume}{26}, \bibinfo{number}{2} (\bibinfo{year}{2024}), \bibinfo{pages}{537--553}.
\newblock
\href{https://doi.org/10.1287/msom.2022.0132}{doi:\nolinkurl{10.1287/msom.2022.0132}}


\bibitem[Chaney et~al\mbox{.}(2018)]%
        {chaney2018confounding}
\bibfield{author}{\bibinfo{person}{Allison~JB Chaney}, \bibinfo{person}{Brandon~M Stewart}, {and} \bibinfo{person}{Barbara~E Engelhardt}.} \bibinfo{year}{2018}\natexlab{}.
\newblock \showarticletitle{How algorithmic confounding in recommendation systems increases homogeneity and decreases utility}. In \bibinfo{booktitle}{\emph{Proceedings of the 12th ACM conference on recommender systems}}. \bibinfo{pages}{224--232}.
\newblock


\bibitem[Chang et~al\mbox{.}(2022)]%
        {chang2022graphrr}
\bibfield{author}{\bibinfo{person}{Yaomin Chang}, \bibinfo{person}{Lin Shu}, \bibinfo{person}{Erxin Du}, \bibinfo{person}{Chuan Chen}, \bibinfo{person}{Ziyang Zhang}, \bibinfo{person}{Zibin Zheng}, \bibinfo{person}{Yuzhao Huang}, {and} \bibinfo{person}{Xingxing Xing}.} \bibinfo{year}{2022}\natexlab{}.
\newblock \showarticletitle{GraphRR: A multiplex Graph based Reciprocal friend Recommender system with applications on online gaming service}.
\newblock \bibinfo{journal}{\emph{Knowledge-Based Systems}}  \bibinfo{volume}{251} (\bibinfo{year}{2022}), \bibinfo{pages}{109187}.
\newblock


\bibitem[Dudík et~al\mbox{.}(2014)]%
        {dudik2014doubly}
\bibfield{author}{\bibinfo{person}{Miroslav Dudík}, \bibinfo{person}{Dumitru Erhan}, \bibinfo{person}{John Langford}, {and} \bibinfo{person}{Lihong Li}.} \bibinfo{year}{2014}\natexlab{}.
\newblock \showarticletitle{Doubly Robust Policy Evaluation and Optimization}.
\newblock \bibinfo{journal}{\emph{Statist. Sci.}} \bibinfo{volume}{29}, \bibinfo{number}{4} (\bibinfo{year}{2014}).
\newblock
\showISSN{0883-4237}
\href{https://doi.org/10.1214/14-sts500}{doi:\nolinkurl{10.1214/14-sts500}}


\bibitem[Edmonds and Karp(1972)]%
        {edmonds1972theoretical}
\bibfield{author}{\bibinfo{person}{Jack Edmonds} {and} \bibinfo{person}{Richard~M Karp}.} \bibinfo{year}{1972}\natexlab{}.
\newblock \showarticletitle{Theoretical improvements in algorithmic efficiency for network flow problems}.
\newblock \bibinfo{journal}{\emph{Journal of the ACM (JACM)}} \bibinfo{volume}{19}, \bibinfo{number}{2} (\bibinfo{year}{1972}), \bibinfo{pages}{248--264}.
\newblock


\bibitem[Gale and Shapley(1962)]%
        {gale1962college}
\bibfield{author}{\bibinfo{person}{David Gale} {and} \bibinfo{person}{Lloyd~S. Shapley}.} \bibinfo{year}{1962}\natexlab{}.
\newblock \showarticletitle{College Admissions and the Stability of Marriage}.
\newblock \bibinfo{journal}{\emph{The American Mathematical Monthly}} \bibinfo{volume}{69}, \bibinfo{number}{1} (\bibinfo{year}{1962}), \bibinfo{pages}{9--15}.
\newblock
\href{https://doi.org/10.2307/2312726}{doi:\nolinkurl{10.2307/2312726}}


\bibitem[Gao et~al\mbox{.}(2024)]%
        {gao2022causal}
\bibfield{author}{\bibinfo{person}{Chen Gao}, \bibinfo{person}{Yu Zheng}, \bibinfo{person}{Wenjie Wang}, \bibinfo{person}{Fuli Feng}, \bibinfo{person}{Xiangnan He}, {and} \bibinfo{person}{Yong Li}.} \bibinfo{year}{2024}\natexlab{}.
\newblock \showarticletitle{Causal inference in recommender systems: A survey and future directions}.
\newblock \bibinfo{journal}{\emph{ACM Transactions on Information Systems}} \bibinfo{volume}{42}, \bibinfo{number}{4} (\bibinfo{year}{2024}), \bibinfo{pages}{1--32}.
\newblock


\bibitem[Jain and Vaish(2024)]%
        {rosenfeld2018fair}
\bibfield{author}{\bibinfo{person}{Pallavi Jain} {and} \bibinfo{person}{Rohit Vaish}.} \bibinfo{year}{2024}\natexlab{}.
\newblock \showarticletitle{Maximizing nash social welfare under two-sided preferences}. In \bibinfo{booktitle}{\emph{Proceedings of the AAAI Conference on Artificial Intelligence}}, Vol.~\bibinfo{volume}{38}. \bibinfo{pages}{9798--9806}.
\newblock


\bibitem[Joachims et~al\mbox{.}(2017)]%
        {joachims2017unbiased}
\bibfield{author}{\bibinfo{person}{Thorsten Joachims}, \bibinfo{person}{Adith Swaminathan}, {and} \bibinfo{person}{Tobias Schnabel}.} \bibinfo{year}{2017}\natexlab{}.
\newblock \showarticletitle{Unbiased Learning-to-Rank with Biased Feedback}. In \bibinfo{booktitle}{\emph{Proceedings of the Tenth ACM International Conference on Web Search and Data Mining}} (Cambridge, United Kingdom) \emph{(\bibinfo{series}{WSDM '17})}. \bibinfo{publisher}{Association for Computing Machinery}, \bibinfo{address}{New York, NY, USA}, \bibinfo{pages}{781–789}.
\newblock
\showISBNx{9781450346757}
\href{https://doi.org/10.1145/3018661.3018699}{doi:\nolinkurl{10.1145/3018661.3018699}}


\bibitem[Joseph et~al\mbox{.}(2016)]%
        {joseph2016rawls}
\bibfield{author}{\bibinfo{person}{Matthew Joseph}, \bibinfo{person}{Michael Kearns}, \bibinfo{person}{Jamie Morgenstern}, \bibinfo{person}{Seth Neel}, {and} \bibinfo{person}{Aaron Roth}.} \bibinfo{year}{2016}\natexlab{}.
\newblock \showarticletitle{Rawlsian fairness for machine learning}.
\newblock \bibinfo{journal}{\emph{arXiv preprint arXiv:1610.09559}} \bibinfo{volume}{1}, \bibinfo{number}{2}, \bibinfo{pages}{19}.
\newblock


\bibitem[Kingma and Ba(2014)]%
        {kingma2014adam}
\bibfield{author}{\bibinfo{person}{Diederik~P. Kingma} {and} \bibinfo{person}{Jimmy Ba}.} \bibinfo{year}{2014}\natexlab{}.
\newblock \showarticletitle{Adam: {A} Method for Stochastic Optimization}.
\newblock \bibinfo{journal}{\emph{arXiv preprint}}  \bibinfo{volume}{arXiv:1412.6980} (\bibinfo{year}{2014}).
\newblock
\urldef\tempurl%
\url{https://arxiv.org/abs/1412.6980}
\showURL{%
\tempurl}


\bibitem[Klimashevskaia et~al\mbox{.}(2022)]%
        {jannach2022popbias}
\bibfield{author}{\bibinfo{person}{Anastasiia Klimashevskaia}, \bibinfo{person}{Mehdi Elahi}, \bibinfo{person}{Dietmar Jannach}, \bibinfo{person}{Christoph Trattner}, {and} \bibinfo{person}{Lars Skj{\ae}rven}.} \bibinfo{year}{2022}\natexlab{}.
\newblock \showarticletitle{Mitigating Popularity Bias in Recommendation: Potential and Limits of Calibration Approaches}. In \bibinfo{booktitle}{\emph{Advances in Bias and Fairness in Information Retrieval}}, \bibfield{editor}{\bibinfo{person}{Ludovico Boratto}, \bibinfo{person}{Stefano Faralli}, \bibinfo{person}{Mirko Marras}, {and} \bibinfo{person}{Giovanni Stilo}} (Eds.). \bibinfo{pages}{82--90}.
\newblock
\showISBNx{978-3-031-09316-6}


\bibitem[Leskovec et~al\mbox{.}(2009)]%
        {leskovec2008community}
\bibfield{author}{\bibinfo{person}{Jure Leskovec}, \bibinfo{person}{Kevin~J Lang}, \bibinfo{person}{Anirban Dasgupta}, {and} \bibinfo{person}{Michael~W Mahoney}.} \bibinfo{year}{2009}\natexlab{}.
\newblock \showarticletitle{Community structure in large networks: Natural cluster sizes and the absence of large well-defined clusters}.
\newblock \bibinfo{journal}{\emph{Internet Mathematics}} \bibinfo{volume}{6}, \bibinfo{number}{1} (\bibinfo{year}{2009}), \bibinfo{pages}{29--123}.
\newblock


\bibitem[Li et~al\mbox{.}(2022)]%
        {li2023stableDR}
\bibfield{author}{\bibinfo{person}{Haoxuan Li}, \bibinfo{person}{Chunyuan Zheng}, {and} \bibinfo{person}{Peng Wu}.} \bibinfo{year}{2022}\natexlab{}.
\newblock \showarticletitle{StableDR: Stabilized doubly robust learning for recommendation on data missing not at random}.
\newblock \bibinfo{journal}{\emph{arXiv preprint arXiv:2205.04701}}.
\newblock


\bibitem[Liu(2023)]%
        {joseph2023twosided}
\bibfield{author}{\bibinfo{person}{Jiqun Liu}.} \bibinfo{year}{2023}\natexlab{}.
\newblock \showarticletitle{Toward a two-sided fairness framework in search and recommendation}. In \bibinfo{booktitle}{\emph{Proceedings of the 2023 Conference on Human Information Interaction and Retrieval}}. \bibinfo{pages}{236--246}.
\newblock


\bibitem[Nakada et~al\mbox{.}(2024)]%
        {nakada2024parallel}
\bibfield{author}{\bibinfo{person}{Kento Nakada}, \bibinfo{person}{Kazuki Kawamura}, {and} \bibinfo{person}{Ryosuke Furukawa}.} \bibinfo{year}{2024}\natexlab{}.
\newblock \showarticletitle{Parallel and Mini-Batch Stable Matching for Large-Scale Reciprocal Recommender Systems}.
\newblock \bibinfo{journal}{\emph{arXiv preprint}}  \bibinfo{volume}{arXiv:2411.19214} (\bibinfo{year}{2024}).
\newblock
\urldef\tempurl%
\url{https://arxiv.org/abs/2411.19214}
\showURL{%
\tempurl}


\bibitem[Neve and Palomares(2019)]%
        {neve2019latent}
\bibfield{author}{\bibinfo{person}{James Neve} {and} \bibinfo{person}{Ivan Palomares}.} \bibinfo{year}{2019}\natexlab{}.
\newblock \showarticletitle{Latent factor models and aggregation operators for collaborative filtering in reciprocal recommender systems}. In \bibinfo{booktitle}{\emph{Proceedings of the 13th ACM conference on recommender systems}}. \bibinfo{pages}{219--227}.
\newblock


\bibitem[Orlin(1997)]%
        {orlin1997polynomial}
\bibfield{author}{\bibinfo{person}{James~B Orlin}.} \bibinfo{year}{1997}\natexlab{}.
\newblock \showarticletitle{A polynomial time primal network simplex algorithm for minimum cost flows}.
\newblock \bibinfo{journal}{\emph{Mathematical Programming}} \bibinfo{volume}{78}, \bibinfo{number}{2} (\bibinfo{year}{1997}), \bibinfo{pages}{109--129}.
\newblock


\bibitem[Palomares et~al\mbox{.}(2021)]%
        {palomares2021}
\bibfield{author}{\bibinfo{person}{Iván Palomares}, \bibinfo{person}{Carlos Porcel}, \bibinfo{person}{Luiz Pizzato}, \bibinfo{person}{Ido Guy}, {and} \bibinfo{person}{Enrique Herrera-Viedma}.} \bibinfo{year}{2021}\natexlab{}.
\newblock \showarticletitle{Reciprocal Recommender Systems: Analysis of state-of-art literature, challenges and opportunities towards social recommendation}.
\newblock \bibinfo{journal}{\emph{Information Fusion}}  \bibinfo{volume}{69} (\bibinfo{year}{2021}), \bibinfo{pages}{103--127}.
\newblock
\showISSN{1566-2535}
\href{https://doi.org/10.1016/j.inffus.2020.12.001}{doi:\nolinkurl{10.1016/j.inffus.2020.12.001}}


\bibitem[Pan et~al\mbox{.}(2021)]%
        {chen2021feedbackloop}
\bibfield{author}{\bibinfo{person}{Weishen Pan}, \bibinfo{person}{Sen Cui}, \bibinfo{person}{Hongyi Wen}, \bibinfo{person}{Kun Chen}, \bibinfo{person}{Changshui Zhang}, {and} \bibinfo{person}{Fei Wang}.} \bibinfo{year}{2021}\natexlab{}.
\newblock \showarticletitle{Correcting the User Feedback-Loop Bias for Recommendation Systems}.
\newblock  (\bibinfo{year}{2021}).
\newblock
\showeprint[arxiv]{2109.06037}~[cs.IR]
\urldef\tempurl%
\url{https://arxiv.org/abs/2109.06037}
\showURL{%
\tempurl}


\bibitem[Pisani(2025)]%
        {pisani2025exposure}
\bibfield{author}{\bibinfo{person}{Andrea Pisani}.} \bibinfo{year}{2025}\natexlab{}.
\newblock \showarticletitle{On the Longitudinal Impact of Exposure Bias in Recommender Systems}. In \bibinfo{booktitle}{\emph{Advances in Information Retrieval: 47th European Conference on Information Retrieval, ECIR 2025, Lucca, Italy, April 6–10, 2025, Proceedings, Part V}} (Lucca, Italy). \bibinfo{pages}{178–183}.
\newblock
\showISBNx{978-3-031-88719-2}
\href{https://doi.org/10.1007/978-3-031-88720-8_29}{doi:\nolinkurl{10.1007/978-3-031-88720-8_29}}


\bibitem[Pizzato et~al\mbox{.}(2013)]%
        {pizzato2013recommending}
\bibfield{author}{\bibinfo{person}{Luiz Pizzato}, \bibinfo{person}{Tomasz Rej}, \bibinfo{person}{Joshua Akehurst}, \bibinfo{person}{Irena Koprinska}, \bibinfo{person}{Kalina Yacef}, {and} \bibinfo{person}{Judy Kay}.} \bibinfo{year}{2013}\natexlab{}.
\newblock \showarticletitle{Recommending people to people: the nature of reciprocal recommenders with a case study in online dating}.
\newblock \bibinfo{journal}{\emph{User Modeling and User-Adapted Interaction}} \bibinfo{volume}{23}, \bibinfo{number}{5} (\bibinfo{year}{2013}), \bibinfo{pages}{447--488}.
\newblock


\bibitem[Potts et~al\mbox{.}(2018)]%
        {potts2018reciprocal}
\bibfield{author}{\bibinfo{person}{Boyd~A. Potts}, \bibinfo{person}{Hassan Khosravi}, \bibinfo{person}{Carl Reidsema}, {and} \bibinfo{person}{Aneesha Bakharia}.} \bibinfo{year}{2018}\natexlab{}.
\newblock \showarticletitle{Reciprocal Peer Recommendation for Learning Purposes}. In \bibinfo{booktitle}{\emph{Proceedings of the 8th International Conference on Learning Analytics \& Knowledge (LAK '18)}}. \bibinfo{pages}{226--235}.
\newblock
\href{https://doi.org/10.1145/3170358.3170400}{doi:\nolinkurl{10.1145/3170358.3170400}}


\bibitem[Roth(1984)]%
        {roth1984evolution}
\bibfield{author}{\bibinfo{person}{Alvin~E Roth}.} \bibinfo{year}{1984}\natexlab{}.
\newblock \showarticletitle{The evolution of the labor market for medical interns and residents: a case study in game theory}.
\newblock \bibinfo{journal}{\emph{Journal of political Economy}} \bibinfo{volume}{92}, \bibinfo{number}{6} (\bibinfo{year}{1984}), \bibinfo{pages}{991--1016}.
\newblock


\bibitem[Schnabel et~al\mbox{.}(2016)]%
        {schnabel2016}
\bibfield{author}{\bibinfo{person}{Tobias Schnabel}, \bibinfo{person}{Adith Swaminathan}, \bibinfo{person}{Ashudeep Singh}, \bibinfo{person}{Navin Chandak}, {and} \bibinfo{person}{Thorsten Joachims}.} \bibinfo{year}{2016}\natexlab{}.
\newblock \showarticletitle{Recommendations as treatments: Debiasing learning and evaluation}. In \bibinfo{booktitle}{\emph{international conference on machine learning}}. PMLR, \bibinfo{pages}{1670--1679}.
\newblock


\bibitem[S{\"u}hr et~al\mbox{.}(2019)]%
        {suhr2019twosided}
\bibfield{author}{\bibinfo{person}{Tom S{\"u}hr}, \bibinfo{person}{Asia~J. Biega}, \bibinfo{person}{Meike Zehlike}, \bibinfo{person}{Krishna~P. Gummadi}, {and} \bibinfo{person}{Abhijnan Chakraborty}.} \bibinfo{year}{2019}\natexlab{}.
\newblock \showarticletitle{Two-Sided Fairness for Repeated Matchings in Two-Sided Markets: A Case Study of a Ride-Hailing Platform}. In \bibinfo{booktitle}{\emph{Proceedings of the 25th ACM SIGKDD International Conference on Knowledge Discovery \& Data Mining (KDD '19)}}. \bibinfo{publisher}{Association for Computing Machinery}, \bibinfo{address}{New York, NY, USA}, \bibinfo{pages}{3082--3092}.
\newblock
\href{https://doi.org/10.1145/3292500.3330793}{doi:\nolinkurl{10.1145/3292500.3330793}}


\bibitem[Swaminathan and Joachims(2015)]%
        {swaminathan2015self}
\bibfield{author}{\bibinfo{person}{Adith Swaminathan} {and} \bibinfo{person}{Thorsten Joachims}.} \bibinfo{year}{2015}\natexlab{}.
\newblock \showarticletitle{Counterfactual Risk Minimization: Learning from Logged Bandit Feedback}. In \bibinfo{booktitle}{\emph{Proceedings of the 32nd International Conference on Machine Learning}} \emph{(\bibinfo{series}{Proceedings of Machine Learning Research}, Vol.~\bibinfo{volume}{37})}, \bibfield{editor}{\bibinfo{person}{Francis Bach} {and} \bibinfo{person}{David Blei}} (Eds.). \bibinfo{publisher}{PMLR}, \bibinfo{pages}{814--823}.
\newblock
\urldef\tempurl%
\url{https://proceedings.mlr.press/v37/swaminathan15.html}
\showURL{%
\tempurl}


\bibitem[Tomita and Yokoyama(2024)]%
        {tomita2024fairrec}
\bibfield{author}{\bibinfo{person}{Yoji Tomita} {and} \bibinfo{person}{Tomohiko Yokoyama}.} \bibinfo{year}{2024}\natexlab{}.
\newblock \showarticletitle{Fair Reciprocal Recommendation in Matching Markets}. In \bibinfo{booktitle}{\emph{Proceedings of the 18th ACM Conference on Recommender Systems (RecSys ’24)}}.
\newblock
\href{https://doi.org/10.1145/3640457.3688130}{doi:\nolinkurl{10.1145/3640457.3688130}}


\bibitem[Tong et~al\mbox{.}(2023)]%
        {gao2024feedback}
\bibfield{author}{\bibinfo{person}{Ding Tong}, \bibinfo{person}{Qifeng Qiao}, \bibinfo{person}{Ting-Po Lee}, \bibinfo{person}{James McInerney}, {and} \bibinfo{person}{Justin Basilico}.} \bibinfo{year}{2023}\natexlab{}.
\newblock \showarticletitle{Navigating the Feedback Loop in Recommender Systems: Insights and Strategies from Industry Practice}. In \bibinfo{booktitle}{\emph{Proceedings of the 17th ACM Conference on Recommender Systems}} (Singapore, Singapore) \emph{(\bibinfo{series}{RecSys '23})}. \bibinfo{pages}{1058–1061}.
\newblock
\showISBNx{9798400702419}
\href{https://doi.org/10.1145/3604915.3610246}{doi:\nolinkurl{10.1145/3604915.3610246}}


\bibitem[Xia et~al\mbox{.}(2015)]%
        {xia2015reciprocal}
\bibfield{author}{\bibinfo{person}{Peng Xia}, \bibinfo{person}{Benyuan Liu}, \bibinfo{person}{Yizhou Sun}, {and} \bibinfo{person}{Cindy Chen}.} \bibinfo{year}{2015}\natexlab{}.
\newblock \showarticletitle{Reciprocal recommendation system for online dating}. In \bibinfo{booktitle}{\emph{Proceedings of the 2015 IEEE/ACM International Conference on Advances in Social Networks Analysis and Mining 2015}}. \bibinfo{pages}{234--241}.
\newblock


\bibitem[Yang et~al\mbox{.}(2024)]%
        {yang2024crrs}
\bibfield{author}{\bibinfo{person}{Chen Yang}, \bibinfo{person}{Sunhao Dai}, \bibinfo{person}{Yupeng Hou}, \bibinfo{person}{Wayne~Xin Zhao}, \bibinfo{person}{Jun Xu}, \bibinfo{person}{Yang Song}, {and} \bibinfo{person}{Hengshu Zhu}.} \bibinfo{year}{2024}\natexlab{}.
\newblock \showarticletitle{Revisiting Reciprocal Recommender Systems: Metrics, Formulation, and Method}. In \bibinfo{booktitle}{\emph{Proceedings of the 30th ACM SIGKDD Conference on Knowledge Discovery and Data Mining}}. \bibinfo{pages}{3714--3723}.
\newblock


\bibitem[Zheng et~al\mbox{.}(2021)]%
        {fang2021dice}
\bibfield{author}{\bibinfo{person}{Yu Zheng}, \bibinfo{person}{Chen Gao}, \bibinfo{person}{Xiang Li}, \bibinfo{person}{Xiangnan He}, \bibinfo{person}{Yong Li}, {and} \bibinfo{person}{Depeng Jin}.} \bibinfo{year}{2021}\natexlab{}.
\newblock \showarticletitle{Disentangling user interest and conformity for recommendation with causal embedding}. In \bibinfo{booktitle}{\emph{Proceedings of the web conference 2021}}. \bibinfo{pages}{2980--2991}.
\newblock


\bibitem[Zhu et~al\mbox{.}(2020)]%
        {zhu2020cjl}
\bibfield{author}{\bibinfo{person}{Ziwei Zhu}, \bibinfo{person}{Yun He}, \bibinfo{person}{Yin Zhang}, {and} \bibinfo{person}{James Caverlee}.} \bibinfo{year}{2020}\natexlab{}.
\newblock \showarticletitle{Unbiased Implicit Recommendation and Propensity Estimation via Combinational Joint Learning}. In \bibinfo{booktitle}{\emph{Proceedings of the 14th ACM Conference on Recommender Systems}} (Virtual Event, Brazil) \emph{(\bibinfo{series}{RecSys '20})}. \bibinfo{pages}{551–556}.
\newblock
\showISBNx{9781450375832}
\href{https://doi.org/10.1145/3383313.3412210}{doi:\nolinkurl{10.1145/3383313.3412210}}


\end{thebibliography}

\end{document}